\documentclass[12pt]{article}
\usepackage{graphicx,lscape}
\usepackage{multirow} % Needed for the ta
\usepackage{epsfig}
\usepackage{overcite}
\usepackage{color}
\newcommand{\epsuv}{$\epsilon_{uv}$}

\begin{document}

\begin{center}

\noindent {\LARGE {\bf Size and interaction dependent solute diffusion in a dilute body centered
cubic solid solution}}

\noindent{\bf Manju Sharma$^1$ and S. Yashonath$^{1,2,\dagger}$}\\

\baselineskip=12pt
\noindent {\it $^{1}$ Solid Sate and Structural Chemistry Unit}\\
\noindent {\it $^{2}$ Center for Condensed Matter Theory}\\ 
\noindent {\it Indian Institute of Science, Bangalore, India - 560 012}\\

\end{center}

\footnotetext {$\dagger$  Also at Jawaharlal Nehru Centre for Advanced 
Scientific Research, Jakkur, Bangalore }

\baselineskip=12pt

\vspace*{1.0cm}
\begin{center}
{\large\bf Abstract}
\end{center}
\noindent

We report results of molecular dynamics simulations to understand the 
role of solute and solute-solvent interaction on solute diffusivity in a
solid solution within a body centered cubic solid when the solute size 
is significantly smaller than the size of the solvent atom. 
Results show that diffusivity is maximum for two specific sizes of the 
solute atom. This is the first time that twin maxima have been found. 
The solute with diffusivity maxima are larger in case of rigid host as 
compared to flexible host. This suggests that the effect of lattice 
vibrations is to decrease the size at which the maximum is seen.
For one of the $\epsilon_{uv}$ where two diffusivity maxima have been 
observed, we have analyzed various properties to understand the 
anomalous diffusion behavior. 
It is characterized by a lower activation energy, lower backscattering 
in the velocity autocorrelation function, lower mean square force, single 
exponential decay of the intermediate scattering function and
monotonic dependence on $k$ of the $\Delta \omega/2Dk^2$ where 
$\Delta \omega$ is the fwhm of the self part of the dynamic structure 
factor. Among the two solute atoms at the anomalous maxima, the solute 
with higher diffusivity has lower activation energy.

\baselineskip=22pt

\section {Introduction}

Diffusion of solute atoms in solids play a significant role in corrosion,
steel hardening, solid batteries, among others. There have been 
innumerable studies of diffusion in close-packed solids. These attempt 
to understand and investigate diffusion in a variety of elements as a 
function of size, temperature, etc. Lee, Ijima and Hirano \cite{Lee} 
investigated diffusion of gallium and indium in $\beta$-titanium.
They studied the temperature dependence and found deviation from the 
Arrhenius behavior and attributed this to phonon-assisted diffusion 
jumps via monovacancies. Further, they found that the activation energy 
is proportional to square of the radius of the diffusing atom.
They attributed this to the predominant influence of size on self diffusion.

Ferro \cite{Ferro} proposed a theory for diffusion in interstitial solid 
solutions of body-centered cubic (b.c.c.) metals. He attempted to evaluate the 
activation energy for interstitial diffusion from the distortion energy necessary 
for the passage of the interstitial atom. He further showed
that the activation energies are in good agreement with experimental data and are related
to the elastic constants. He also studied the dependence of activation energy on diameter
of the interstitial atom.

Most studies in the literature investigate diffusion of solute when the solute size
relative to solvent are not very small. That is, the solute-solvent size ratio is
between 0.8-1.5.  Hood \cite{Hood} analyzed published tracer diffusion data in Pb and $\alpha$-Zr 
at 0.6$T_m$ where $T_m$ is the
melting temperature. They found a striking correlation between diffusivity $D$ and radius
of the metallic element, the diffusant. Published data were fitted to yield a 
relationship between activation enthalpy and radius of the tracer element. They found 
the activation enthalpy were lowest for Cu and Ni in Pb (around 8 kcal/mol) and these 
also had the highest diffusivities. Further, the radius of these were sufficiently 
small to avoid overlap of these atoms with the ion core of Pb. This is one study
where the solute is small relative to the solvent size. 

%In this present study, we have investigated the size dependence of diffusivity of 
%solute atoms over a wide range of size in a body centered cubic solid. We have studied the 
%solute diffusion as a function of the size and solute-solvent interaction.  
 
Here we report a detailed molecular dynamics study of dependence of self diffusivity of solute in
a body-centred cubic (b.c.c.) matrix, the solvent. The solute-solvent size ratio $\sigma_u/\sigma_v$  is 
varied over 0.06-0.44 while keeping the solvent size the same throughout. This corresponds
to solute sizes that are comparable or smaller than the void and neck sizes present in the
b.c.c. solid. A single maximum or 
two maxima in self diffusivity are seen as a function of the solute size depending on the
strength the interaction between solute and solvent. Related properties such as the velocity
autocorrelation function, intermediate scattering function and other functions yield 
interesting insights into the nature of motion of the solute. 

\section{Methods}

\subsection {Intermolecular potential}

\begin{figure}
\begin{center}
{\includegraphics*[width=6cm]{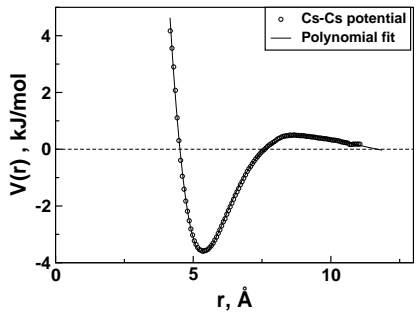}}
\caption{ Caesium atom interatomic potential. Full line represent the potential data from reference\cite{Shyu, yasho_bcc}
and broken line is the potential obtained from the polynomial fit.} 
\label{Cs_pot}
\end{center}
\end{figure}

Solvent atoms are arranged in a body centered cubic (b.c.c.) arrangement and the 
solute atoms are placed at the center of tetrahedral voids chosen randomly. The 
stable structure of the solid is crucially determined by the interatomic 
potential \cite{Parrinello_Rahman_solid}. Therefore, the 
interatomic potential given by Shyu \cite{Shyu} for caesium has been 
employed here. This potential was used by Yashonath and Rao for Monte Carlo studies
in solids \cite{yasho_bcc}. As alkali metal atoms have a stable body-centred cubic structure, the
use of this potential ensures a stable b.c.c. host solid. The potential
shown in Figure \ref{Cs_pot} is fitted to a  polynomial of the form given in Eq. \ref{polynom}. 

\begin{equation}
V(r)=\frac{a_0}{r^9}-\frac{a_1}{r^4}-\frac{a_2}{r^2}+\frac{a_3}{r^{11}}-\frac{a_4}{r^6}-\frac{a_5}{r^8}
\label{polynom}
\end{equation}

%The values of the coefficients are: $a_0$=-1.58831$\times$$10^{9}$kJ/mol\AA$^{-9}$, 
%$a_1$=-108804.0kJ/mol\AA$^{-4}$, $a_2$=427.427.0kJ/mol\AA$^{-2}$, 
%$a_3$=4.39526$\times$$10^{9}$kJ/mol\AA$^{-11}$, $a_4$=9.20694$\times$$10^{6}$kJ/mol\AA$^{-6}$, 
%$a_5$=-4.5013$\times$$10^{8}$kJ/mol\AA$^{-8}$. The polynomial and the coefficients were used 
%for solvent-solvent interaction. 
Note that the solvent-solvent interaction 
energy changes from positive to negative around 4.5\AA\ which gives us an approximate
diameter for solvent atom. The solute atoms are smaller in size than the
solvent atoms. The solute-solute as well as solute-solvent interactions are 
modeled in terms of the Lennard-Jones interactions. The interaction parameters are 
reported in Table\ref{intern_par}. The solute-solvent Lennard Jones diameter is chosen using 
the rule $\sigma_{uv} = \sigma_{uu} + 0.7$\AA \cite{pradipliq, Vash_rahman, Parrinello_Vashishta}.

\begin{equation}
 \phi_{\alpha\beta} = 4\epsilon_{\alpha\beta}\left[\left(\frac{\sigma_{\alpha\beta}}
{r_{\alpha\beta}}\right)^{12}
- \left(\frac{\sigma_{\alpha\beta}}{r_{\alpha\beta}}\right)^6\right] 
\label{LJ_eqn}
\end{equation}

The total interaction energy of the system is a sum of solvent-solvent, 
$U_{vv}$, solvent-solute, $U_{uv}$ and solute-solute, $U_{vv}$ interaction energy. 
 
\begin{equation}
U_{tot} = U_{vv} + U_{uv} + U_{uu}
\label{Totenr}
\end{equation}

\section {Computational Details}

Simulations have been performed in the microcanonical ensemble at a reduced density, $\rho^*$ of 1.09 and 5488
solvent atoms with 588 solute atoms. The mass of solvent and solute species is 132.9 and 40.0amu 
respectively. The solvent-solvent interaction is the 
same in all the simulations while the solute size is varied over the 
range 0.3-2.0\AA\ which corresponds to solute-solvent size ratio of 0.06-0.44. 
All simulation runs are performed with Verlet leapfrog scheme 
using DLPOLY \cite{dlpoly} at 60K. A timestep of 2fs yielded relative standard 
deviation in total energy of the order of 10$^{-5}$. Cut-off radius is 17\AA. The system is 
equilibrated for 1ns and positions, velocities and forces of the solute atoms are stored at an interval 
of 250fs for 2ns. 

\begin{landscape}
\begin{table}
\begin{center}
\caption{\label{table1} Caesium and Lennard-Jones interaction parameters employed in the present study.}
~\\
\begin{tabular}{|c|c|c|c|c|c|c|}\hline
 {Type of} & \multicolumn{6}{c|}{Interatomic Potential Parameters} \\\cline{2-7}
 {Interaction} & \multicolumn{3}{c|}{$\sigma$, \AA} & \multicolumn{3}{c|}{$\epsilon,
 kJ/mol$}  \\\cline{1-7}

 {uu} & \multicolumn{3}{c|} {0.3 - 2.0} & \multicolumn{3}{c|} {0.4} \\\cline{1-7}
 {uv} & \multicolumn{3}{c|} {1.0 - 2.7} & \multicolumn{3}{c|} {3.0} \\\cline{1-7}

\multirow{3}{*}{vv} & $a_0$ & $a_1$ & $a_2$ & $a_3$ & $a_4$ & $a_5$  \\

& kJ/mol \AA$^{-9}$ & kJ/mol \AA$^{-4}$ & kJ/mol \AA$^{-2}$ & kJ/mol \AA$^{-11}$ & kJ/mol \AA$^{-6}$ & kJ/mol \AA$^{-8}$  \\\cline{2-7}
& -1.5883 $\times 10^{9}$ & -1.0880 $\times 10^{5}$ & 427.427 & 4.3953 $\times 10^{9}$ & 9.2064 $\times 10^{6}$ & 4.5013 $\times 10^{9}$  \\\cline{1-7}
\end{tabular}
\label{intern_par}
\end{center}
\end{table}
\end{landscape}

\section {Results and Discussion}

\begin{figure}
\begin{center}
{\includegraphics*[width=6cm]{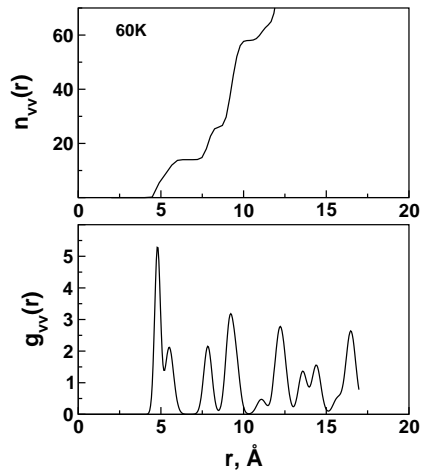}}
\caption{Solvent-solvent radial distribution function for host system and number of
nearest solvent atoms for a given solvent atom. The results are for system at 60K.} 
\label{rdf_n_bcc}
\end{center}
\end{figure}

The solvent-solvent radial distribution function at 60K is shown in Figure \ref{rdf_n_bcc}. 
The integrated value as a function of $r$, 
$n_{vv}(r)=\int_0^r \rho g_{vv}(r^{\prime})4\pi r^{\prime 2} dr^{\prime}$ 
is also reported in the figure. The shoulder to the first peak is typical of b.c.c. solids.
From the radial distribution function as well as
the number of neighbours within a radius $r$ seen above, it is evident that the 
structure of the solid is body-centred cubic. The number of neighbours in the first
few peaks have 8, 6, 12, 32 and 6, etc neighbouring solvent species. These suggest that the
structure is a b.c.c. solid. The peak positions in the solvent-solvent rdf are at
4.80:5.53:7.85:9.24:11.03:12.22.  The ratio of the square ($r^2$) of the peak positions is 
23.04:30.58:61.62:85.38:121.66:149.33. Dividing this by the value for the first peak we get
1:1.327:2.67:3.71:5.28:6.48 which is in the proportion 
1:4/3:8/3:11/3:4:16/3:19/3, etc. expected for a b.c.c. solid.\cite{Parrinello_Rahman_solid}  
We see that in the 
present study the peaks corresponding to 11/3 and 4 are merger giving only a single peak at
3.71. But the overall peak positions and their intensities are consistent with the
b.c.c. structure. 

\begin{figure}
\begin{center}
{\includegraphics*[width=6cm]{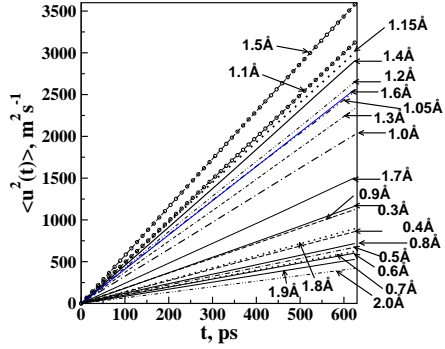}}
\caption{Mean square displacement as a function of time of the solutes of various sizes
diffusing in a bcc solid.}
\label{msd_bcc}
\end{center}
\end{figure}

\begin{figure}
\begin{center}
{\includegraphics*[width=6cm]{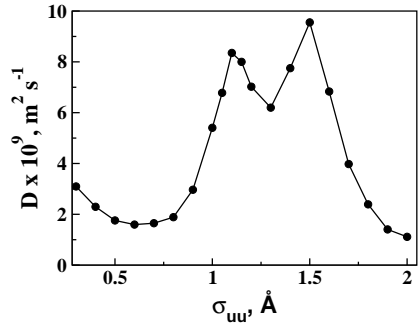}}
\caption{Self diffusivity, $D$ as a function of size of solute, $\sigma_{uu}$. There 
are diffusivity maxima at 1.1\AA\ and 1.5\AA.} 
\label{Dsigma_bcc}
\end{center}
\end{figure}

Figure \ref{msd_bcc} shows the mean square displacement (MSD) for various solute 
sizes. We see that the curves are straight suggesting
good statistics. The diffusivities are obtained from the slope of MSD data using Einstein's
relationship and these are plotted as a function of solute size in Figure \ref{Dsigma_bcc}. 
The diffusivity of solute initially decreases with increase in size of solute. Then 
there is a gradual increase in diffusivity with increase 
in solute size as can be seen from size 0.7\AA\ onwards with a 
diffusivity maximum at 1.1\AA. On further increase in solute diameter, there is a decrease in 
diffusivity which is followed by an increase in diffusivity once again for size 1.4\AA\ and 
a second diffusivity 
maximum is observed for solute size 1.5\AA. The diffusivity of 1.5\AA\ solute is 
slightly higher than that of 1.1\AA. The diffusivity of solutes 
larger than 1.5\AA\ sharply decrease with size.

Diffusion of a relatively small solute within the solvent made up of large-sized atoms
occurs through jumps from one void to another neighbouring void. Two neighbouring voids
are connected through a narrower region which is referred to as neck. Passage through
the neck indeed forms the bottleneck for diffusion. The motion of the solvent
atoms is around their equilibrium positions this leads to a distribution of 
neck diameters, f($r_n$) instead of a unique value for the neck diameter. 
It may be necessary for the solvent 
atoms to move away from their equilibrium position to permit the solute to pass through
the bottleneck. This leads to activation energy barriers for large solutes. Large here
means solutes whose diameter is larger than the neck diameter of the void interconnecting
two neighbouring voids.

We have computed the strain energy in b.c.c. solid for different sized solutes.
A small system consisting of 432 solvent atoms in a b.c.c. solid with 5 solute atoms has been
simulated at 5K to compute the strain energy. Position coordinates were stored at an interval of 250fs for 500ps.
Two sets of simulations were carried out, one in which the solvent atoms were not included in the
molecular dynamics simulations (and therefore they were rigid) and another in which the solvent
atoms were included in the integration along with the solute atoms. 
The average solute-solvent interaction energy is calculated in the case of rigid as well as
flexible solvent runs. The difference in solute-solvent interaction energy in flexible and rigid host
is the strain energy. Table \ref{strain_E} reports the solute-solvent interaction of
various solute atoms and the strain energy. We note that the strain energy is generally
small unless the solute diameter is large which is to be expected. Even for the largest
sized solute of 1.8\AA\ which is well beyond the size for which the diffusivity maximum was
found in our simulations, the stain energy is about 10\%\ of the total solute-solvent
energy. Thus, in the regimes which are relevant to the present study and the regime where
the diffusivity maximum is seen the contribution from the strain energy is not more 
than 16\%.  Most studies in the literature including those that were discussed in the introduction
mainly concern large solutes where strain energy is important but for the present work,
it plays only a secondary role, if at all. From this it is clear that although strain
energy might be responsible for {\em decrease} in the diffusivity after the diffusivity
maximum, diffusivity maximum does not have its origin in the strain energy.

\begin{table}
\caption{Strain energy of solute atoms in linear and anomalous regimes.}
\begin{center}
\begin{tabular}{ccccc}\hline
$\sigma_{uu}$, & Regime & $U_{uv}$(flexible host)& $U_{uv}$(rigid host)& Strain energy\\\hline
\AA & & kJ/mol & kJ/mol & kJ/mol \\ \hline\hline
0.5&Linear&-2.7430&-2.7660& 0.023\\
0.6&Linear&-2.8334&-2.8132&-0.0202\\
0.9&Linear&-2.4911&-2.4704&-0.0207\\
1.1&Anomalous&-4.4226&-3.5568&-0.8658\\
1.3&Linear&-5.7127&-5.3544&-0.3583\\
1.5&Anomalous&-7.3439&-6.6290&-0.7149\\
1.8&Linear&-11.2414&-9.9361&-1.3053\\\hline
\end{tabular}
\label{strain_E}
\end{center}
\end{table}

Before we analyse the results that might lead to an understanding of the diffusivity maximum,
we discuss the influence of the solute-solvent interaction energy on the diffusivity maximum.

\subsection {Two distinct diffusivity maxima}

%This is the first time that twin size dependent diffusivity maxima is seen. 

There are previous reports of the existence of diffusivity maximum as a function of diffusants
confined to other condensed matter phases such as porous solids, liquids, amorphous solids, etc \cite{yashosanti94b, padma, pradipliq, manju_glass}. 
However, this is the first time that such a maximum has been reported for body-centred cubic 
close-packed solid. These studies have invariably reported a single diffusivity maximum 
as a function of the size of the diffusant. But here we find two maxima which has not been
reported previously. It is therefore of considerable interest to investigate when and how
such twin maxima are seen. 

It is well known that solute-solvent interaction plays an important role in giving rise to 
anomalous diffusivity maximum. For example, it has been demonstrated unambiguously that the
the size dependent diffusivity maximum of solute disappears in 
the absence of attractive interaction between the solute and solvent 
medium.\cite{yashosanti94b, SE_manju}
Further, it has been shown that the diffusivity maximum disappears when the magnitude of the
diffusant-medium (in the present case solute-solvent) interaction is small relative
the kinetic energy, $k_BT$. We have therefore carried out simulations with different values of
$\epsilon_{uv}$, the solute-solvent interaction strength. 
%In addition, we have 
%chosen three different host systems (body centered cubic solid (rigid and flexible), face 
%centered cubic solid (flexible) and dense fluid (flexible)). 
Simulations were carried out on a smaller system consisting of 686 solvent and 68 solute atoms at 70K.
%Face centered cubic
%solid has a reduced density 0.933 with 500 solvent and 50 solute atoms ($\epsilon_{vv}=1.2kJ/mol,
%\epsilon_{uv}=3.0kJ/mol, \epsilon_{uu}=0.4kJ/mol, \sigma_{vv}=4.5\AA$). 
%The f.c.c. solid melted at 160K to obtain dense fluid and simulations 
%in dense fluid were done at this temperature. 
The position coordinates were stored once every 2ps.
% and at 250fs for f.c.c. solid and dense fluid.

\begin{figure}
\begin{center}
{\includegraphics*[width=6cm]{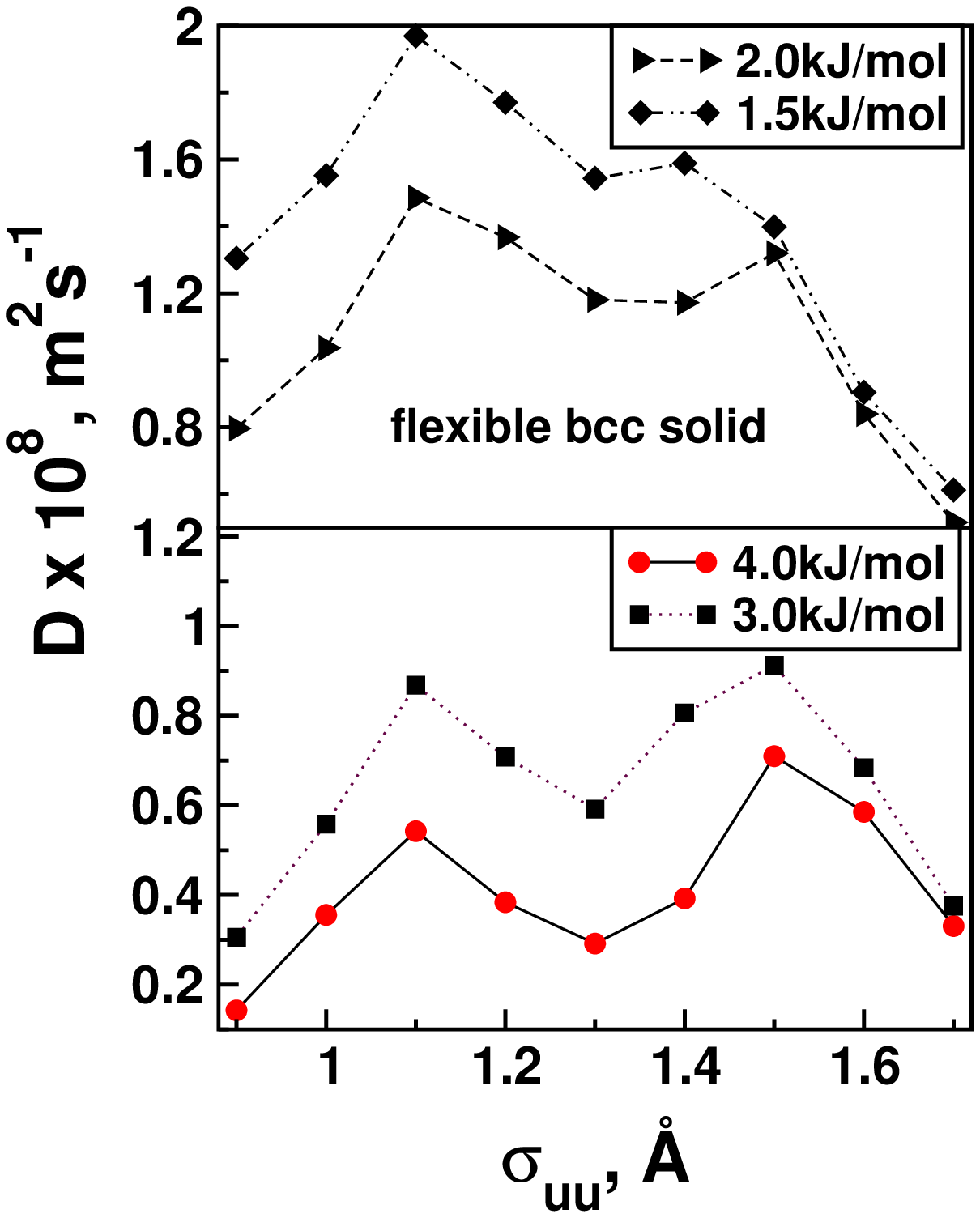}}\hspace*{0.5cm}
{\includegraphics*[width=6cm]{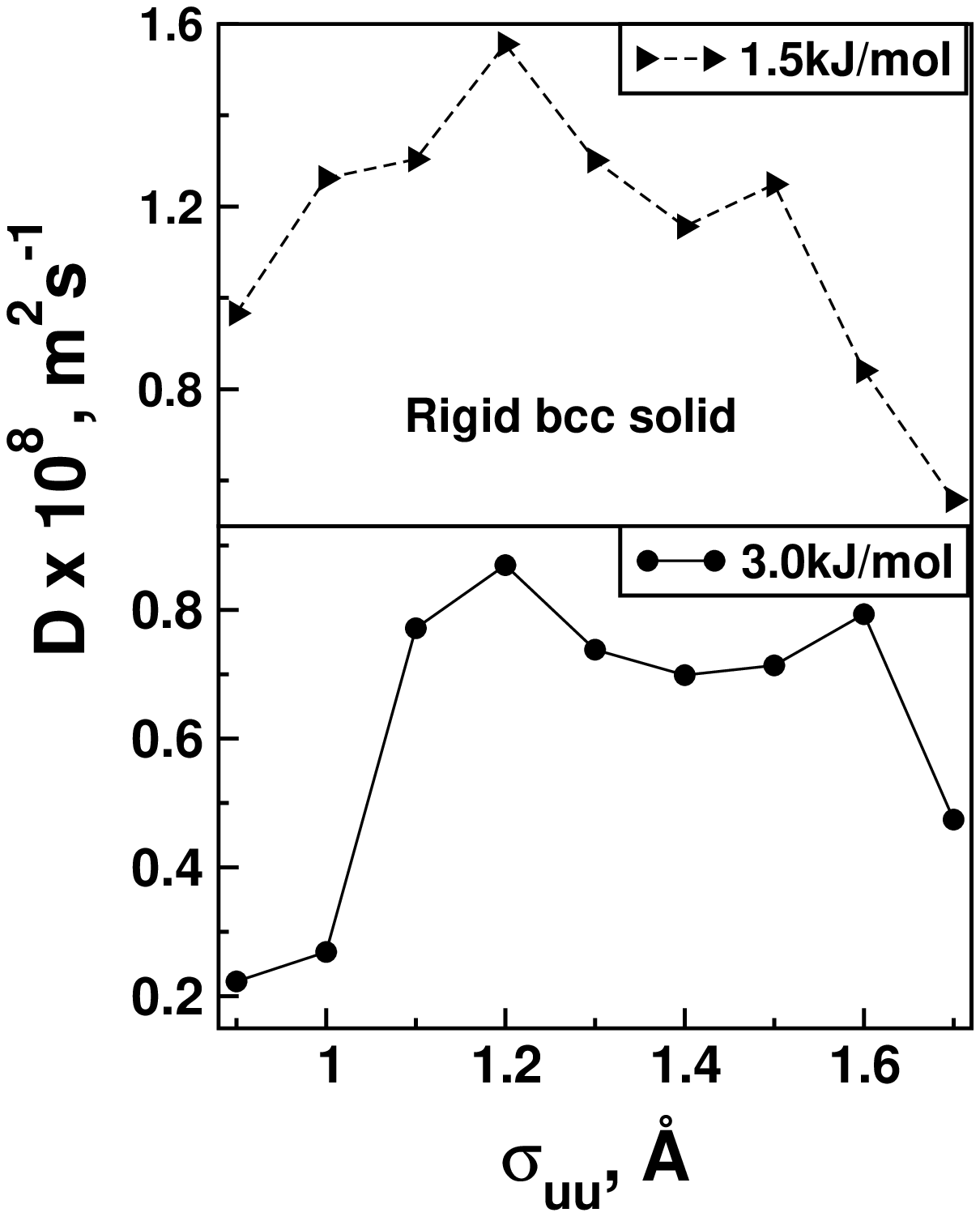}}\\\vspace*{0.6cm}
{\includegraphics*[width=8cm]{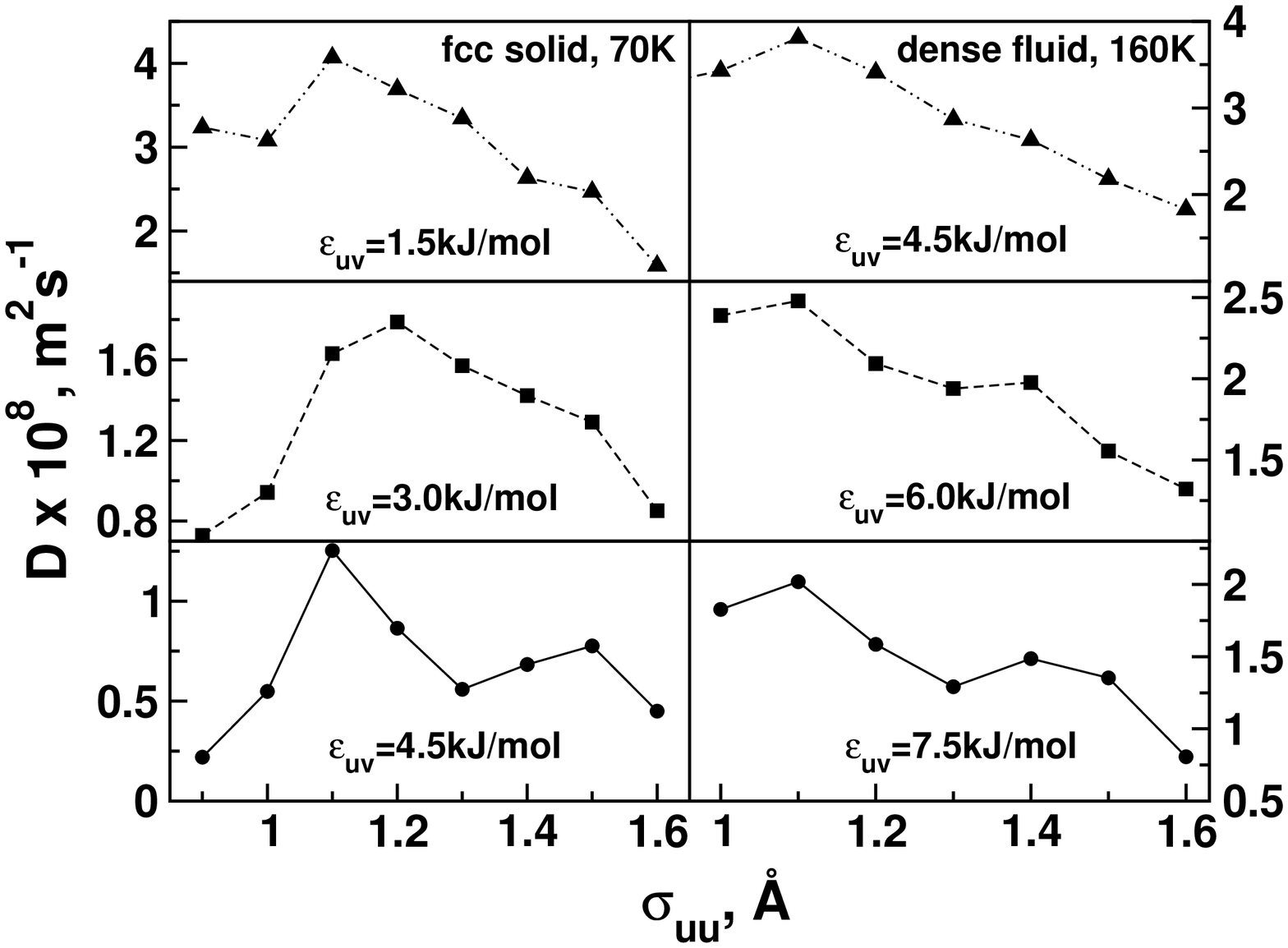}}
\caption{Size dependent diffusivity of solute atoms as a function of solute-solvent interaction, 
$\epsilon_{uv}$ in rigid and flexible bcc solid, flexible fcc solid and dense fluid.}
\label{anoml_check}
\end{center}
\end{figure}

The diffusivity of solutes as a function of solute size is plotted in 
Figure \ref{anoml_check}. In case of flexible bcc host, $\epsilon_{uv}$ has been varied over
the range 1.5-4.0 kJ/mol. Although there are two maxima (at 1.1 and 1.4\AA) for \epsuv\ = 1.5 kJ/mol,
the prominent maximum is for 1.1\AA. With increase in interaction strength, the second maxima
at 1.5\AA\ gradually becomes the predominant maxima. We have carried out, simulations with
flexible b.c.c. solid as well as rigid (when the solvent atoms were not included in MD integration).
The effect of increase in \epsuv\ on D-$\sigma$ is same. However, the effect of rigid b.c.c.
lattice is to shift the $\sigma$ at which maxima is seen to the right by 0.1\AA : that is,
the maxima are now seen at 1.2 and 1.6\AA. In order to see if this effect 
of \epsuv\ always leads 
to appearance of a second maxima which further gains in strength, we carried out simulations
on face-centred cubic (f.c.c.) solid solvent as well. The simulations have been carried out on
at a reduced density of 0.933 with 500 solvent and 50 solute atoms  and ($\epsilon_{vv}$=1.2kJ/mol,
$\epsilon_{uv}$=3.0kJ/mol, $\epsilon_{uu}$=0.4kJ/mol, $\sigma_{vv}$=4.5\AA). 
The melting point of this f.c.c. solid is 160K. Coordinates were accumulated
every 250fs to obtain various properties. These results are also shown in Figure \ref{anoml_check}.
We see that a single maximum is seen for low \epsuv\ while for \epsuv\ = 4.5 kJ/mol, there are
two diffusivity maxima. The positions of these are precisely the same as the b.c.c. solid :
1.1 and 1.5\AA. We also carried out simulations of liquid at 160K when the f.c.c. solid melts with same
parameters as the f.c.c. solid. These results are also shown in Figure \ref{anoml_check}.
Note that the for \epsuv\ = 4.5 kJ/mol there is single maximum. This is expected since at high
temperatures, there is significant dynamical disorder. By 7.5 kJ/mol, we see that another 
maxima is developing around 1.4\AA. The principal results of these set of simulations are :
(i) at sufficiently high \epsuv\ two diffusivity maxima instead of the previously observed
single maximum has been seen. (ii) this result is valid irrespective of the solvent structure (b.c.c.
or f.c.c.) (iii) in the fluid phase, it is seen that two maxima are seen only at values of
\epsuv\ that are higher than required for the solid phase. Further investigations are necessary
to understand the role of \epsuv\ as well as distribution of neck diameters, f($r_n$)
on diffusivity maxima.

\subsection {Related aspects of diffusivity maximum}

In order to understand the motion that leads to anomalous diffusivity maximum 
for certain solute sizes, we have obtained several other
properties. The properties of anomalous regime solute sizes are compared
with the properties of solutes which are not part of the maximum. These solutes are those belonging to
the region where self diffusivity decreases with increase in solute diameter. This regime is referred
to as the linear regime.

\begin{figure}
\begin{center}
{\includegraphics*[width=8cm]{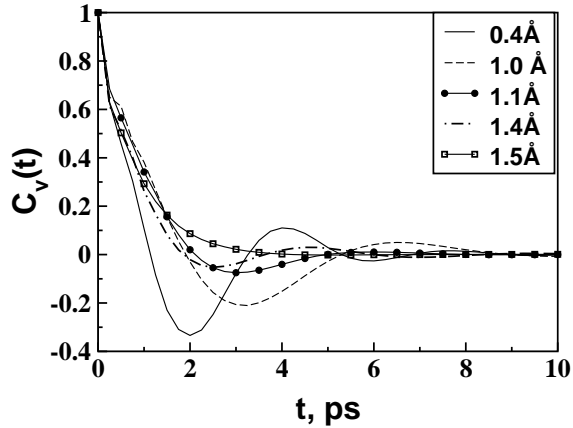}}
\caption{Velocity autocorrelation plot of solute atoms in linear (0.4, 1.0, 1.4)\AA\ 
and anomalous (1.1, 1.5)\AA\ regime.} 
\label{vcf_bcc}
\end{center}
\end{figure}

Figure \ref{vcf_bcc} displays the velocity autocorrelation function (VACF) for solute atoms
in linear and anomalous regime. Solute from the linear regime shows an oscillatory VACF 
whereas solute from anomalous regime exhibits a smoothly decaying VACF without any backscattering. 
We attribute the presence of backscattering to the fact that the solute from linear regime
encounters an energy barrier. Until it has enough energy to overcome the barrier, it performs 
oscillatory motion. Later, we will see how such a barrier arises. The VACF of 1.5\AA\ decays faster in
the initial time period and also exhibits a smoother decay as compared to 1.1\AA\ solute.

\begin{figure}
\begin{center}
{\includegraphics*[width=8cm]{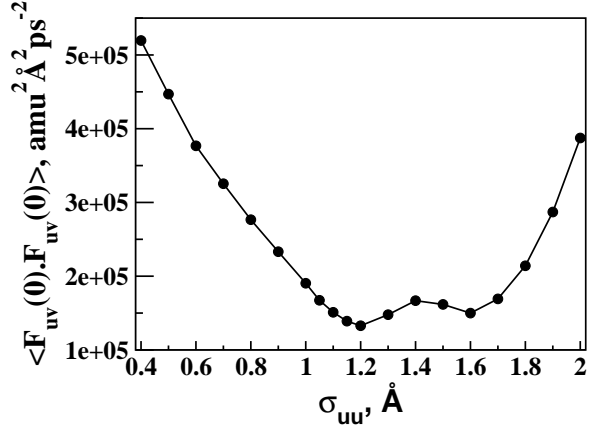}}
\caption{Average mean square force acting on the solute atoms due to the solvent atoms.} 
\label{bcc_Fgh}
\end{center}
\end{figure}

The average mean square force acting on the solute atoms due to the solvent atoms is
shown in Figure \ref{bcc_Fgh}. The anomalous regime solute atoms experience lower average
mean square force as compared to linear regime solute atoms. When the size of solute is very small 
as compared to the bottleneck diameter, it feels a large net force due to the neighboring solvent atoms 
in the neck region. The linear regime solute is thus bound and has a lower diffusivity.
The anomalous regime solute has a diameter comparable to the neck diameter. For this reason,
during its passage through the neck of the solute from anomalous regime, the centre of mass 
of the solute coincides  with the centre of the bottleneck. By symmetry, the force exerted by
the solvent atoms in a given direction is equal and opposite to the force exerted along 
diagonally opposite direction. This results in a mutual cancellation of forces leading to
lower net force. The average mean square force on 1.1\AA\ solute is smaller than solute of size 1.5\AA\ 
though diffusivity of 1.5\AA\ is higher  than 1.1\AA. This result is important since
it suggests for the first time that there are factors other than mean square force that 
influence the diffusivity. We will see what these may be.

\pagebreak

\begin{figure}
\begin{center}
{\includegraphics*[width=8cm]{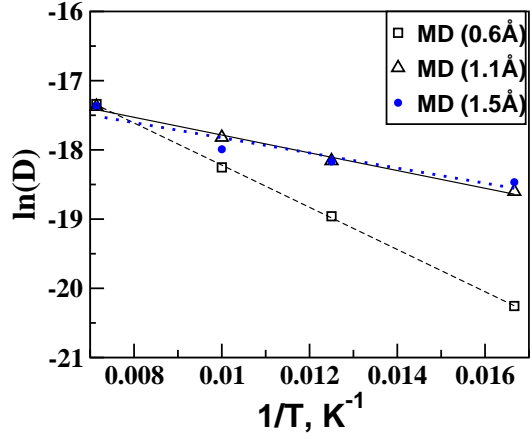}}
\caption{Arrhenius plot for solute atoms in linear and anomalous regime obtained from
diffusivities at 60, 80, 100 and 140K.} 
\label{bcc_arrh}
\end{center}
\end{figure}

The diffusivity of a solute depends crucially on its activation energy.
Therefore, we have computed $D$ at several temperatures.
Arrhenius plot of linear and anomalous regime solute 
atoms from the diffusivities at four different temperatures (60, 80, 100 and 140K) for
different solute sizes is shown in Figure \ref{bcc_arrh}. 

The activation energy of the solute atoms obtained from the slope of Arrhenius plot are
listed in Table \ref{Eact}. The activation energy of linear regime solute atoms is larger than anomalous
regime solute atoms. Further, activation energy of 1.5\AA\ solute is lower than 1.1\AA\ which
is consistent with the observed higher diffusivity of 1.5\AA\ solute as compared to 1.1\AA.
Thus, it appears that the activation energy is responsible for the observed differences in
self diffusivity as a function of the size. In particular, the difference in the 
self diffusivity of the solutes located at the two maxima can be explained in terms of the difference 
in the activation energy for these two solute sizes.

\begin{table}
\caption {Activation Energy of solute atoms in linear and anomalous regimes.}
\begin{center}
\begin{tabular}{{c}{c}{c}}\hline
$\sigma_{uu}$, \AA & Regime & $E_{act}$, kJ/mol\\\hline\hline
0.6&linear&2.5333\\
1.1&anomalous&1.0659\\
1.5&anomalous&0.9096\\\hline
\end{tabular}
\label{Eact}
\end{center}
\end{table}

\subsection {Physical picture of motion of the solute}

More detailed behaviour of the motion of the solute can be gleaned from the wavenumber 
dependence of various properties. The self part of the intermediate scattering function, 
$F_s$(k,t) has been computed from the molecular dynamics data.

\begin{figure}
\begin{center}
{\includegraphics*[width=8cm]{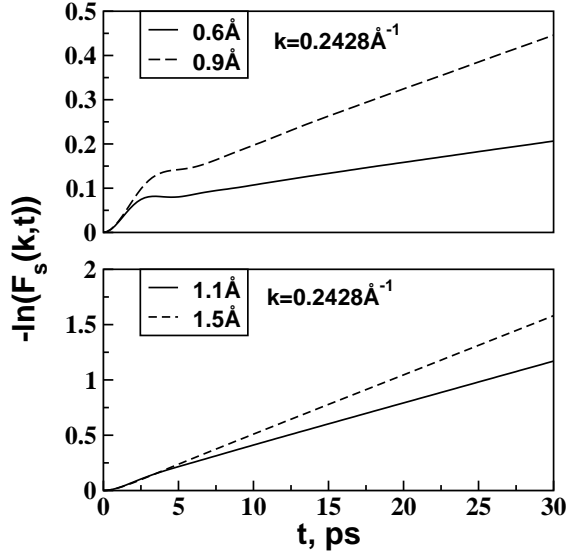}}
\caption{-ln($F_s$(k,t)) as a function of time for linear (0.6, 0.9)\AA\ and 
anomalous (1.1, 1.5)\AA\ regime solute atoms for motion within b.c.c. solid.}
\label{logfskt}
\end{center}
\end{figure}

The natural logarithm of $F_s$(k,t) as a function of time for 
solute atoms in both the linear and anomalous regimes is reported in Figure \ref{logfskt}. 
The inverse of the slope 
of -ln($F_s$(k,t)) as a function of time gives the relaxation time. In case of linear regime solute 
atoms, there are two slopes corresponding to two relaxation times whereas anomalous regime solute atoms 
show only one slope and one relaxation time. This suggests that the solute atom from the linear regime 
performs two distinct type of motion while anomalous regime impurity atom performs only one type of
motion. 
                                                    
The physical picture regarding the motion performed by solutes in these regimes 
is similar to that proposed by Singwi and Sj\"olander \cite{Singwi} for solute motion in water. 
Water molecules diffuse much more slowly in water than the solute and this is similar to the
present situation where the solvent molecules are slow while the solute motion is fast. The model
due to Singwi and Sj\"olander envisages that the solute performs an oscillatory motion for a short
period of time, $\tau_1$ before it performs a diffusive motion on the time scale $\tau_2$. This
leads to biexponential decay and applies to linear regime solute. We suggest that the 
linear regime solute performs oscillatory motion initially when it is confined in the solvent shell for
a given time. Once it overcomes the energy barrier to move past the solvent shell it performs
motion with relaxation time $\tau_2$. The anomalous regime solute does not feel the energy barrier 
at the solvent shell and thus finds the region within the solvent shell and region outside the solvent
shell similar. It, thus, sees a homogeneous solvent rather than two distinct, heterogeneous regions - one inside
the solvent shell and another outside - seen by the solute from linear regime. 
The relaxation times of different solute atoms are reported in Table \ref{relax_time}.

\begin{table}
\caption{Relaxation times of solute atoms in linear and anomalous regimes for wavevector 0.2428\AA.}
\begin{center}
\begin{tabular}{{c}{c}{c}{c}}\hline
$\sigma_{uu}$, \AA & Regime & $\tau_1$, ps &  $\tau_2$, ps\\\hline\hline
0.6&Linear&31.25&218.44\\
0.9&Linear&23.53&83.33\\
1.1&Anomalous&26.00&\\
1.5&Anomalous&19.23&\\\hline
\end{tabular}
\label{relax_time}
\end{center}
\end{table}

The relaxation time for anomalous regime solute atom is less than the linear regime solute atom. Further, 
the relaxation of 1.5\AA\ solute is faster than 1.1\AA. This is consistent with both the lower
activation energy and the relative magnitudes of self diffusivities.  

The Fourier transformation of self part of the intermediate scattering function gives the dynamic structure
factor, $S_s$(k,$\omega$). In the hydrodynamic limit, full width at half maximum (fwhm) $\Delta\omega$(k) 
of dynamic structure factor is 2D$k^2$. The ratio $\Delta\omega(k)/2Dk^2$ provides an idea of
the $k$ dependence of self diffusivity. This ratio, $\Delta(k)$ is shown in Eq. \ref{ratio_delta}

\begin{equation}
\Delta(k) = \frac{\Delta\omega(k)}{2Dk^2}
\label{ratio_delta}
\end{equation}

\begin{figure}
\begin{center}
{\includegraphics*[width=8cm]{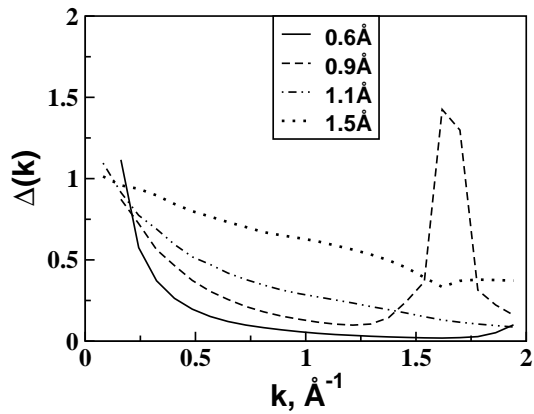}}
\caption{$\Delta$(k) as a function of wavevector, k for linear and anomalous regime solute atoms.}
\label{deltaomega}
\end{center}
\end{figure}

We have obtained the $\Delta$(k) for different wavenumbers for the solute atoms in 
linear and anomalous regimes using the width of dynamic structure factor and is shown in 
Figure \ref{deltaomega}. The width, $\Delta$(k) shows oscillatory behavior for linear regime solute atoms 
and nearly monotonic decay of $\Delta$(k) for anomalous regime solute atoms. The minimum seen 
for $k$ around 1.2\AA$^{-1}$ for 0.9\AA\ solute arises from the slowing down of the self diffusivity
on these length scales. $k$ = 1.2\AA$^{-1}$ approximately corresponds to the position where the first 
neighbour shell is located. These are in good agreement with the results for liquid argon
at low and high densities by Nijboer and Rahman \cite{rahman66} as well as Levesque and Verlet \cite{verlet70}
as well as the discussions in Boon and Yip \cite{boon_yip}. The anomalous regime solute does not experience any 
barrier to exit from the first solvent shell and therefore shows a nearly smooth decay of $\Delta$(k) with
wavenumber.

\section{Conclusions}

The present study suggests that there is a maximum in self diffusivity for solutes diffusing within
the interstitial space provided by a body-centred cubic solid. The maxima are seen when the 
solute/solvent size ratio is in the range 0.25-0.33. We report, for the first time, the existence
of more than one maximum in self diffusivity as a function of the size of the solute. 
We show that two maxima are seen when the solute-solvent interaction strength is large. 
It is seen that two maxima are also seen in face-centred cubic arrangement as well as 
in liquids. For the latter, two maxima are seen when the solute-solvent interaction strength
is higher relative to diffusion in solids. Further, we show that the relative heights of 
the two maxima are determined by the activation energies. We emphasize that the present study
does not study the regime of large solutes where strain energy becomes important.
In previous studies, it was thought that distribution of bottleneck diameter alone had a role
in determining the diffusivity maximum. The present study suggests that
in addition to f($r_n$), solute-solvent interaction strength also influences
the observed size dependence of self diffusivity on diameter of the solute.

\noindent
{\em Acknowledgment} : Authors wish to thank Department of Science and
Technology, New Delhi for financial support in carrying out this work.
Authors also acknowledge C.S.I.R., New Delhi for a research fellowship to 
M.S.

\bibliographystyle{plain}  % Use the "jpc" BibTeX style for formatting the Bibliography
\bibliography{manju_bcc}

\end{document}